\documentclass[runningheads]{llncs}
\usepackage[T1]{fontenc}
\usepackage{graphicx,verbatim}
\usepackage[]{hyperref}
\usepackage{booktabs}
\usepackage{multirow}
\begin{document}
\title{Interactive Segmentation and Report Generation for CT Images }
%

\author{Yannian Gu\textsuperscript{1,*}, Wenhui Lei\textsuperscript{1,*}, Hanyu Chen\textsuperscript{2}, Xiaofan Zhang\textsuperscript{1,\dag}, Shaoting Zhang\textsuperscript{1}}  
\authorrunning{Anonymized Author et al.}
\institute{
\textsuperscript{1}Shanghai Jiao Tong University, China \\
\textsuperscript{2}China Medical University, China \\
\textsuperscript{*}These authors contributed equally. \\
\textsuperscript{\dag}Corresponding author: \email{xiaofan.zhang@sjtu.edu.cn}
}

\maketitle              
\begin{abstract}
Automated CT report generation plays a crucial role in improving diagnostic accuracy and clinical workflow efficiency. 
However, existing methods lack interpretability and impede patient-clinician understanding, while their static nature restricts radiologists from dynamically adjusting assessments during image review.
Inspired by interactive segmentation techniques, we propose a novel interactive framework for 3D lesion morphology reporting that seamlessly generates segmentation masks with comprehensive attribute descriptions, enabling clinicians to generate detailed lesion profiles for enhanced diagnostic assessment.
To our best knowledge, we are the first to integrate the interactive segmentation and structured reports in 3D CT medical images.
Experimental results across 15 lesion types demonstrate the effectiveness of our approach in providing a more comprehensive and reliable reporting system for lesion segmentation and capturing.
The source code will be made publicly available following paper acceptance.


\end{abstract}

\section{Introduction}

CT (Computed Tomography) serves as an essential diagnostic tool, with radiological reports serving as the primary medium for communicating diagnostic findings to clinicians~\cite{lai2024pixel,wang2024data}. 
In recent years, CT report generation has gained significant attention, with advancements focusing on automating the clinical text generation process based on image analysis results~\cite{dayarathna2024deep,hamamci2024ct2rep,hamamci2024generatect,nakaura2024preliminary}. 
Researchers have developed various approaches to generate reports from CT images, include traditional vision-based methods~\cite{hou2024energy}, vision-language models~\cite{blankemeier2024merlin,yin-etal-2025-kia}, and other methods~\cite{li2024kargen,xiang2024gmod}.
However, current CT report generation methods \textbf{lack interpretability} and impede patient-clinician understanding, while offering no interactivity~\cite{luo2024rexplain}. 
This produces reports that fail to explain findings clearly or adapt to new clinical cases, resulting in communication gaps and incomplete assessments.

To address these limitations, interactive segmentation techniques have shown promising capabilities in related medical imaging analysis. 
Models like the Segment Anything Model (SAM)~\cite{kirillov2023segment,ravi2024sam} enable clinicians to dynamically guide the segmentation process through intuitive interactions, providing immediate visual feedback and greater control over results. 
These techniques~\cite{huang2024segment,ma2024segment,mazurowski2023segment} successfully balance automation with expert input, allowing for precise refinements while maintaining workflow efficiency.
However, current interactive segmentation methods lack integrated report generation capabilities, forcing radiologists to manually convert visual findings into structured clinical reports.

In this work, we introduce the first interactive framework for lesion morphology reporting, integrating both intuitive visual segmentation and quantitative textual attribute description for a comprehensive lesion characterization. 
Our system enables radiologists to generate detailed clinical reports proposed by Lei et al.~\cite{lei2025data}, through minimal point-based interactions during image review, significantly reducing interpretation time while preserving diagnostic precision. 
By simultaneously generating accurate segmentation results with structured reports, our approach provides improved explanatory capabilities, creating direct visual-textual correspondence between report descriptions and anatomical features in CT images. 
Additionally, this integrated framework demonstrates considerable potential for facilitating multi-modal medical dataset annotation.

Technically, our approach delivers two key innovations: (1) feature-space clustering-based point refinement  that amplifies the impact of expert interactions, and (2) inter-task feature synergy that enables bidirectional information flow between segmentation and attribute description generation. 
The interactive nature of our system significantly enhances zero-shot performance, allowing radiologists to provide real-time feedback that helps the model adapt to novel lesion types not encountered during training.
Our contributions are as follows:
\begin{itemize}
    \item We propose the interactive framework for joint lesion segmentation and attribute description that integrates point refinement and feature synergy, enabling comprehensive lesion profiling with minimal radiologist interaction;
    \item Our approach significantly improves clinical workflow efficiency by reducing interpretation time while enhancing diagnostic communication through direct visual-textual correspondence between reports and segmentations;
    \item Extensive experiments demonstrate superior performance in both segmentation accuracy and attribute description quality, including robust zero-shot capabilities for novel lesion types.
\end{itemize}

\section{Methodology}

\begin{figure}[t]
    \centering
    \includegraphics[width=0.9\linewidth]{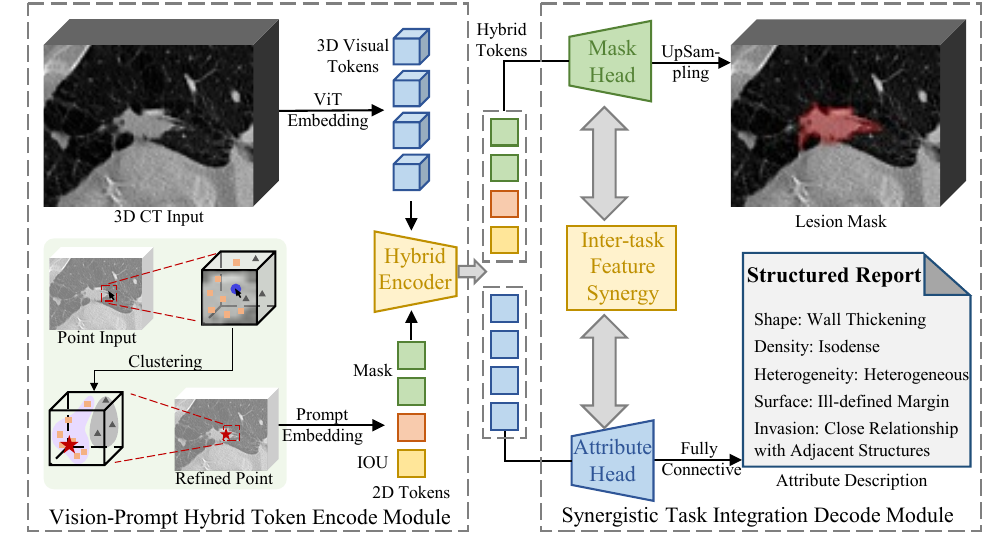}
    \caption{\textbf{The model architecture}. Our framework processes 3D CT images and user-provided points to simultaneously generate lesion segmentation masks and structured attribute descriptions. The model integrates visual tokens, point tokens, initial mask tokens, and initial IOU tokens into a hybrid encoding module, which forms the foundation for both visual and textual outputs. Key innovations include a clustering-based point refinement technique that optimizes the input points through clustering centers, and an inter-task feature synergy mechanism that enhances the performance of both segmentation and attribute description tasks concurrently.}
    \label{pipeline}
\end{figure}

Our proposed interactive framework enables radiologists to generate detailed lesion profiles through minimal point interactions. 
As shown in Fig.~\ref{pipeline}, the model consists of two main components: the \textit{Vision-Prompt Hybrid Token Encode} module and the \textit{Synergistic Task Integration Decode} module. 
Below, we describe each module in detail.

\subsection{Vision-Prompt Hybrid Token Encode Module}
This module follows a dual-branch architecture, with image tokens extracted by the \textit{vision encoder} and prompt tokens derived from the \textit{prompt encoder}. 
Notably, to enhance the accuracy and reliability of user clicks, the prompt encoder integrates a \underline{feature-space clustering-based point refinement} method.
The refined prompt tokens are then combined with the image tokens in the \textit{hybrid token encoder}, producing hybrid tokens that drive subsequent processing.

\textbf{Vision encoder $\mathrm{E}_I$} adopts a ViT-style~\cite{dosovitskiy2020image} architecture, consisting of a standard ViT with primary local window attention and several interleaved global attention layers.
This design produces isotropic feature maps with consistent feature dimensions.
The extracted image features $\mathbf{I} = \mathrm{E}_I(\mathcal{I})$, where $\mathcal{I}$ denotes the input image, are down-sampled for computational efficiency.

\textbf{Prompt encoder $\mathrm{E}_P$} transforms point coordinates using positional embeddings, generating distinct representations based on point labels.
This process converts sparse point prompts into embeddings that guide the model's attention toward specific regions of interest.
To compensate for the randomness and potential errors in individual point placements and improve interaction efficiency, we incorporate feature-space clustering that extends the influence of each point beyond its immediate location.
For each user click $p$, we define a local feature window $\mathbf N(p)$ and obtain features from the image features $\mathbf I$:
\begin{equation}
    \mathbf I_p=\{\mathbf I(q)|q\in\mathbf N(p)\}.
\end{equation}
Applying $K$-means clustering, we partition $\mathbf I_p$ into $k$ clusters and select the point closest to each centroid $\mu_i$ as guidance:
\begin{equation}
    q^*_i=\arg\min_{q\in C_i}\|\mathbf I(q)-\mu_i\|, i=1,\cdots,k.
\end{equation}
These selected points provide structured feedback, enabling the model to refine predictions across semantically related regions rather than isolated pixels.
The final point prompt tokens are obtained by $\mathbf P = \mathrm E_P(q^*)$.

\textbf{Hybrid token encoder $\mathrm E_H$} combines image tokens $\mathbf I$, prompt tokens $\mathbf P$, mask tokens $\mathbf{M}$, and $\mathbf{IOU}$ tokens via self-attention and cross-attention mechanisms, producing the hybrid token $\mathbf{Z} = \mathrm{E}_H(\mathbf{I}, \mathbf{P}, \mathbf{M}, \mathbf{IOU})$. 
Self-attention aggregates prompt context while cross-attention enables interaction with image information. 
The mask tokens $\mathbf{M}$ help initialize and guide the lesion segmentation process, while the $\mathbf{IOU}$ tokens provide an initial metric for optimizing the segmentation through the intersection-over-union score. 

\subsection{Synergistic Task Integration Decode Module}
After obtaining the hybrid tokens, we use the \textit{mask head} and the \textit{attribute head} to generate the segmentation masks and attribute descriptions, respectively.
Recognizing the intrinsic relationship between these tasks, we introduce the \underline{\textit{inter-task feature synergy}} mechanism that enables bidirectional information flow, allowing both tasks to mutually enhance each other's performance.

\textbf{Mask head $\mathrm H_{seg}$}: Initial mask tokens $\mathbf M_0$ interact with the hybrid token $\mathbf Z$ through multiple Transformer layers to capture task-specific spatial and semantic features, formulated as $\mathbf M = \mathrm H_{seg}(\mathbf M_0, \mathbf Z)$.
This contextual refinement process allows the mask tokens to encode detailed structural information about the target regions.
The extracted mask tokens $\mathbf M$ are subsequently upsampled to enhance spatial resolution:
$\hat{\mathbf{M}}=\mathrm{Upsample}(\mathbf M)$, producing the final segmentation mask with precise boundary delineation.

\textbf{Attribute head $\mathrm H_{attr}$}: This component predicts the attributes of the segmented regions by processing the shared hybrid token $\mathbf Z$.
First, $\mathbf Z$ is passed through residual block to capture essential features. 
Next, a self-attention mechanism refines the attribute representations, yielding $\mathbf{A} = \mathrm H_{attr}(\mathbf Z)$.
By modeling both local and global dependencies, this architecture enhances the accuracy of attribute predictions.
Finally, a classification layer with sigmoid activation generates the multi-label outputs, $\hat{\mathbf Y}=\mathrm{FCN}(\mathbf A)$.

\textbf{Inter-task feature synergy}: Medical image segmentation and attribute prediction are inherently complementary: morphological features provide strong indicators for pathological attributes, while attribute knowledge aids in precise boundary delineation. 
To leverage this natural synergy, we enable bidirectional feature exchange between the segmentation features $\mathbf M$ and attribute features $\mathbf A$ by projecting them into a common latent space:
\begin{equation}
    \mathbf Z_{seg} = \mathrm{Attn}(\mathbf M, \mathrm{Proj}(\mathbf A)), \quad
    \mathbf Z_{attr} = \mathrm{Attn}(\mathbf A, \mathrm{Proj}(\mathbf M)),
\end{equation}
where $\mathrm{Attn}(\cdot,\cdot)$ represents the cross-attention mechanism that facilitates task-specific features by leveraging complementary information between tasks.
The projection operation $\mathrm{Proj}(\cdot,\cdot)$ ensures that the features from both tasks are aligned in the same feature space.
This bidirectional interaction improves segmentation accuracy and enhances attribute prediction through mutual guidance.

\subsection{Training Strategy}
Our training strategy addresses two key challenges: (1) generating accurate segmentation masks $\hat{\mathbf{M}}$ and (2) predicting correct multi-label attributes $\hat{\mathbf{Y}}$. 
We employ a simulation-based approach to mimic expert guidance and a composite loss function to optimize both objectives simultaneously.

\textbf{Expert-provided Click Simulation}: Expert annotations provide crucial feedback by highlighting misclassified regions and guiding the model to improve boundary delineation and feature representation, but manual corrections are costly.
To emulate expert feedback during training, we identify misclassified regions by computing error maps between the predicted mask $\mathbf{M}_{pred}$ and the ground truth $\mathbf{M}_{gt}$:
\begin{equation}
    \mathbf M_{fn} = \mathbf M_{gt} \land \neg \mathbf M_{pred}, \quad \mathbf M_{fp} = \neg \mathbf M_{gt} \land \mathbf M_{pred}.
\end{equation}
False negatives $\mathbf M_{fn}$ represent missed features, while false positives $\mathbf M_{fp}$ show incorrect predictions. We randomly select a point from these regions as a simulated expert click, guiding the model's attention to improve performance in subsequent iterations.


\textbf{Dual-objective Loss Function}: We optimize our model using a composite loss function that balances segmentation accuracy and attribute classification: $\mathcal L_{total}=\mathcal L_{seg}+\lambda\mathcal L_{attr}$, where $\lambda$ is an adaptive weighting parameter.
For the segmentation component, we employ Dice loss to measure overlap between predicted $\hat{\mathbf M}$ and ground truth masks $\mathbf M_{gt}$:
\begin{equation}
    \mathcal L_{seg}=1-\frac{2\sum_i(\hat m(i)\cdot m_{gt}(i))}{\sum_i\hat m(i)+\sum_i m_{gt}(i)+\epsilon},
\end{equation}
where $\hat m(i)$ and $m_{gt}(i)$ represent the predicted and ground-truth values at voxel $i$, and $\epsilon$ is a small constant to avoid division by zero.
For attribute prediction, we use categorical cross-entropy loss:
\begin{equation}
    \mathcal L_{attr}=-\frac{1}{N}\sum_{n=1}^N\sum_{c=1}^C y_c(n)\log \hat{y}_c(n),
\end{equation}
where $\hat{\mathbf{Y}} = \{\hat{y}_1, \hat{y}_2, \cdots, \hat{y}_C\}$ represents predicted probabilities across $C$ classes, and $\mathbf{Y}{gt}$ contains one-hot encoded ground-truth labels.

\section{Experiments}
In this section, we present a comprehensive evaluation of our approach through a series of experiments designed to assess both segmentation performance and structured report generation capabilities.

\begin{table}[t]
\centering
\begin{minipage}[t]{0.5\textwidth}
\centering
\caption{Templates of Structured Lesion Reports. This table summarizes the five key radiological attributes used to characterize lesions, which represents a distinct aspect of lesion morphology and appearance.}
\footnotesize
\begin{tabular}{cc}
\toprule
Attribute     & Description                                                                                                                                        \\ \midrule
Shape         & \begin{tabular}[c]{@{}c@{}}``Round-like'', ``Irregular'',\\ ``Wall thickening'',\\ ``Punctate, nodular''\end{tabular}                                      \\ \midrule
Invasion      & \begin{tabular}[c]{@{}c@{}}``No close relationship \\ with surrounding structures'',\\ ``Close relationship \\ with adjacent structures''\end{tabular} \\ \midrule
Density       & \begin{tabular}[c]{@{}c@{}}``Hypodense'', ``Isodense'',\\ ``Hyperdence''\end{tabular}                                                                    \\ \midrule
Heterogeneity & \begin{tabular}[c]{@{}c@{}}``Homogeneous'',\\ ``Heterogeneous''\end{tabular}\\ \midrule
Surface       & \begin{tabular}[c]{@{}c@{}}``Well-defined margin'', \\ ``Ill-defined margin''\end{tabular} \\
\bottomrule
\end{tabular}
\label{report detaild}
\end{minipage}
\hfill 
\begin{minipage}[t]{0.45\textwidth}
\centering
\caption{Segmentation Performance. Our proposed method is compared against state-of-the-art segmentation approaches and ablation variants of our model.}
\footnotesize
\begin{tabular}{cccc}
\toprule
& Methods                                  & DSC $\uparrow$      & HD95 $\downarrow$      \\ \midrule
\multirow{6}{*}{test} &
 Baseline(UNet)~\cite{ronneberger2015u}     & 0.733               & 5.586                     \\
& SegMamba~\cite{xing2024segmamba}   & 0.747               & 5.601                    \\
& SAM-Med3D~\cite{wang2023sam} & 0.752               & 5.398                   \\ \cline{2-4}
& Vanilla & 0.758 & 5.020\\
& Point Refinemet & 0.764  & 5.122   \\
& Feature Synergy & 0.766  & 4.928  \\
& Ours & \textbf{0.794} &  \textbf{4.303}\\ 
\midrule
\midrule
\multirow{6}{*}{\begin{tabular}[c]{@{}l@{}}Zero-\\ Shot\end{tabular}} &
 Baseline(UNet) & 0.712               & 2.758 \\
& SegMamba & 0.712               & 2.695     \\
& SAM-Med3D & 0.718               & 2.169     \\\cline{2-4}
& Vanilla & 0.726 & 2.079\\
& Point Refinemet & 0.731 & 2.018\\
& Feature Synergy & 0.727 & 2.113\\
& Ours & \textbf{0.735} & \textbf{2.001}\\
\bottomrule
\end{tabular}
\label{segment table}
\end{minipage}
\end{table}

\subsection{Dataset Description}
We assemble 1535 CT scans and masks collected from public and private sources: KiTS23~\cite{heller2023kits21}: kidney tumor and cyst (489 scans); MSD~\cite{antonelli2022medical}: colon tumor (126 scans), liver tumor (303 scans), lung tumor (96 scans), pancreas tumor (216 scans), pancreas cyst (65 scans); private data collected at *** hospital: liver cyst (30 scans), gallbladder cancer (30 scans), gallstones (30 scans), esophageal cancer (30 scans), gastric cancer (30 scans), kidney stone (30 scans), bladder cancer (30 scans), and bone metastasis (30 scans). All these scans are annotated with structured lesion reports by four radiologists.
The gallstone and liver cyst are treated as zero-shot test cases, meaning they are excluded from the training phase.
The remaining data is divided into training ($60\%$), validation ($20\%$), and test ($20\%$) sets.

\textbf{Structured Report}: We adopt a structured lesion report template, as proposed in~\cite{lei2025data}.
Specifically, each lesion comes with a corresponding structured textual report, including \textit{shape}, \textit{invasion}, \textit{density}, \textit{heterogeneity}, and \textit{surface}, with multiple possible categories for each. Further details can be found in Table~\ref{report detaild}.

\textbf{Data Preprocessing}: In the preprocessing phase, we first locate the largest lesion in each label file to determine its center.
This center is then used to crop the CT volume, ensuring that the region of interest (ROI) is tightly focused on the lesion.
By isolating the lesion and eliminating irrelevant background, the model can focus on the lesion's key features, ultimately improving both the efficiency and accuracy of the learning process.

\begin{table}[t]
\centering
\footnotesize
\renewcommand{\arraystretch}{1.2}
\caption{Performance comparison of different models on lesion structured report generation across multiple features, with accuracy as our evaluation metric.} 
\begin{tabular}{ccccccccc}
\toprule
& & Shape & Density & Invasion & Surface & Invasion & Average \\ \midrule
\multirow{6}{*}{test} & Baseline(CNN) & 0.582 & 0.562 & 0.626 & 0.721 & 0.716 & 0.642 \\
& CT-CLIP~\cite{hamamci2024developing} &  0.681         &     0.525        &      0.769        &   0.759          &   0.746           &    0.696                                 &                        \\
& M3D~\cite{bai2024m3d}   &  0.445            &        0.141     &   0.488           &         0.760    &   0.265           &       0.420      \\ \cline{2-8} 
& Vanilla & 0.680 &	0.605 &	0.769 &	0.796 &	0.752 &	0.720 \\
& Point Refinement & 0.680 & \textbf{0.656} & 0.782 & 0.807 & \textbf{0.762} & 0.738  \\
& Feature Synergy &  0.712 & 0.623 & 0.784 & \textbf{0.850} & 0.756 & 0.745 \\
& Ours & \textbf{0.745} & 0.626 &	\textbf{0.833} &	\textbf{0.850} &	0.741 &	\textbf{0.759} 
      \\ \midrule\midrule
\multirow{6}{*}{\begin{tabular}[c]{@{}l@{}}Zero-\\ Shot\end{tabular}} & Baseline(CNN) & 0.283 & 0.367 & 0.400 & 0.167 & 0.800 & 0.403 \\
& CT-CLIP & 0.283 & 0.500 & 0.317 & 0.000 & \textbf{1.000} & 0.420  \\
& M3D & 0.267 & 0.500 & \textbf{0.700} & 0.000 & 0.000 & 0.293 \\ \cline{2-8} 
& Vanilla &  0.283 & 0.500 & 0.317 & 0.317 & \textbf{1.000} & 0.483 \\
& Point Refinement & 0.283 & 0.483 & 0.317 & \textbf{0.600} & \textbf{1.000} & 0.537 \\
& Feature Synergy  &   0.283 & 0.500 & 0.417 & 0.500 & \textbf{1.000} & 0.540 \\
& Ours & \textbf{0.300} & \textbf{0.600} & 0.483 & 0.533 & \textbf{1.000} & \textbf{0.583} \\ \bottomrule
\end{tabular}
\label{report results}
\end{table}

\subsection{Comparision with the SOTA Methods}
Our evaluation is divided into two key components—segmentation performance and structured attribute description.
All compared methods are either trained or fine-tuned on our datasets to ensure a fair comparison.
\textbf{Ablation Setup}: We evaluate four model variants to isolate component contributions: (1) Vanilla (backbone only), (2) Point Refinement (backbone + feature-space clustering point refinement), (3) Feature Synergy (backbone + Inter-task feature synergy), and (4) our full model combining both enhancements with the backbone.

\textbf{Segmentation Performance}: We compare our model against three leading segmentation methods: \textit{UNet}~\cite{ronneberger2015u} (the well-established baseline in medical image segmentation), \textit{SegMamba}~\cite{xing2024segmamba} (which leverages state space models for sequence modeling in volumetric data), and \textit{SAM-Med3D}~\cite{wang2023sam} (a medical adaptation of the Segment Anything Model with 3D capabilities).
Performance is evaluated using two metrics: Dice Similarity Coefficient (\textit{DSC}), which measures volumetric overlap, and $95\%$ Hausdorff Distance (\textit{HD95}), which captures boundary precision.

As demonstrated in Table~\ref{segment table}, our approach outperforms the compared methods in both standard test cases and more challenging zero-shot scenarios.
These results underscore our method's robustness in handling anatomical variability while maintaining precise boundary delineation.
Specifically, the Vanilla model's strong performance confirms the synergy between segmentation and attribute prediction tasks.
The Point Refinement enhancement demonstrates how focusing on diagnostically relevant regions improves results, while Feature Synergy validates our approach of connecting spatial and semantic features.

\textbf{Structured Attribute Description}: Structured attribute description can be framed both as a multi-label classification task, where each lesion has multiple associated attributes, and as a visual-language task, where the model answers questions about the lesion's characteristics or aligns the image with corresponding text reports.
To evaluate our approach, we compare it against methods from both perspectives with methods \textit{CNN model}, \textit{M3D}~\cite{bai2024m3d}, and \textit{CT-CLIP}~\cite{hamamci2024developing}.
This allows us to benchmark our method against both specialized VLMs and traditional visual models in the medical vision domain.


As seen in Table~\ref{report results}, our method outperforms others in attribute prediction. 
The M3D VQA-based approach struggles with complex features (0.000 accuracy for zero-shot surface characteristics), while the CNN baseline falters with lesion boundaries (accuracy drops from 0.642 to 0.403 in zero-shot). 
CT-CLIP performs well in standard tests but struggles with unseen lesions and morphological features in zero-shot scenarios. 
Our combined segmentation-attribute approach ensures precise spatial-semantic mapping, yielding superior performance.

\subsection{Interactive Framework Performance}
\begin{figure}
    \centering
    \includegraphics[width=1\linewidth]{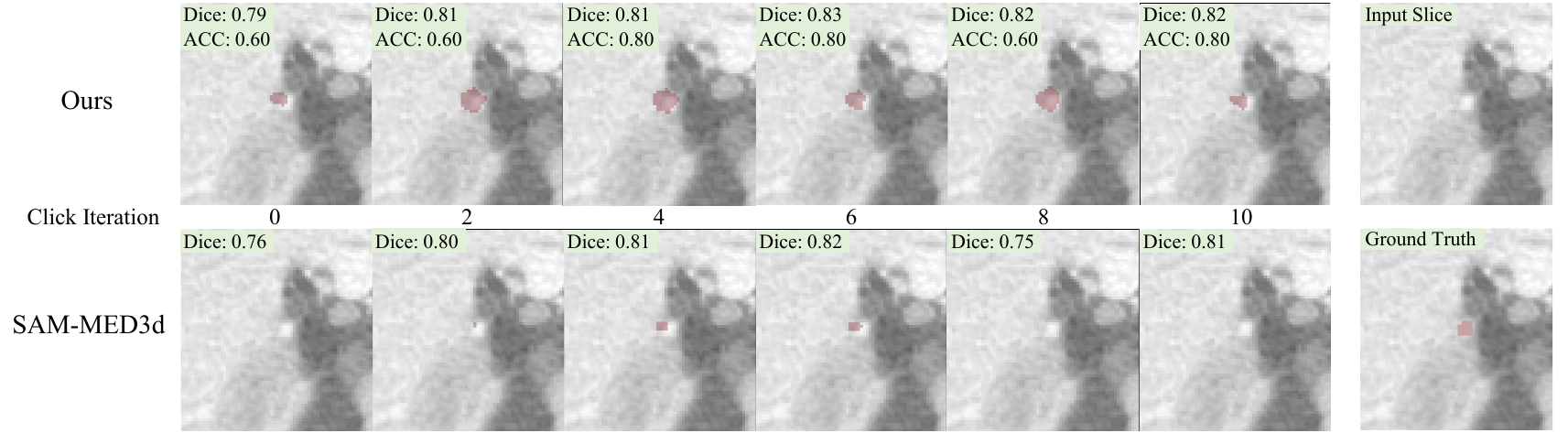}
    \caption{Qualitative comparison between our method and SAM-Med3D, showing segmentation progression across click iterations. Red overlays indicate masks.}
    \label{visualization}
\end{figure}

To evaluate our interactive framework, we compare against SAM-Med3D, focusing on zero-shot cases.
Fig.~\ref{visualization} illustrates qualitative results across click iterations.
Our method achieves better boundary delineation and lesion characterization than SAM-Med3D, with progressive improvements as click iterations increase, confirming its ability to refine reports interactively based on radiologist input.

\section{Conclusion}
We propose an interactive framework for lesion morphology reporting that integrates visual segmentation with structured attribute description. 
Our approach enhances segmentation accuracy, improves attribute description precision, strengthens zero-shot performance, and allows radiologists to refine reports interactively.
Experiments demonstrate that our method outperforms state-of-the-art approaches, offering a more flexible and accurate solution.

%
%
%
\bibliographystyle{splncs04}
\bibliography{reference}
\end{document}